\documentclass{article}
\usepackage{spconf,amsmath,graphicx}
\usepackage{amsfonts}
\usepackage{tabularx}
\usepackage{xcolor}
\usepackage{multirow}
\usepackage{multicol}
\usepackage{adjustbox}
\usepackage{float}
\usepackage{tikz, pgfplots}
\usepackage{ctable}
\usepackage{placeins}
\usepackage{url}
\def\myname#1{\gdef\@myname{{\em #1}\\}}

\title{PromptASR for contextualized ASR with controllable style}
\name{Xiaoyu Yang, Wei Kang, Zengwei Yao, Yifan Yang, Liyong Guo, Fangjun Kuang, Long Lin, Daniel Povey}
\address{Xiaomi Corp. Beijing, China\\
\footnotesize{\{yangxiaoyu6, dpovey\}@xiaomi.com}}
\begin{document}
%
\maketitle
\begin{abstract}
Prompts are crucial to large language models as they provide context information such as topic or logical relationships. Inspired by this, we propose PromptASR, a framework that integrates prompts in end-to-end automatic speech recognition (E2E ASR) systems to achieve contextualized ASR with controllable style of transcriptions. Specifically, a dedicated text encoder encodes the text prompts and the encodings are injected into the speech encoder by cross-attending the features from two modalities. 
When using the ground truth text from preceding utterances as content prompt, the proposed system achieves 21.9\% and 6.8\% relative word error rate reductions on a book reading dataset and an in-house dataset compared to a baseline ASR system. The system can also take word-level biasing lists as prompt to improve recognition accuracy on rare words.
An additional style prompt can be given to the text encoder and guide the ASR system to output different styles of transcriptions.
The code is available at icefall\footnote{https://github.com/k2-fsa/icefall}.
\end{abstract}

\begin{keywords}
Contextualized ASR, Prompts, Transducer, 
\end{keywords}

\vspace{-0.5em}
\section{Introduction}
\label{sec:introduction}
\vspace{-0.3em}
External text information is commonly used to improve E2E ASR systems. Traditional approaches use external language models trained on the text corpora and re-rank the n-best hypotheses~\cite{ASRrescoring, TonyGPT} of ASR systems or perform shallow fusion~\cite{shallowfusion} to modify the posterior predicted by the ASR system. 

Recently, various methods have been proposed to utilize contextual information to improve the accuracy of speech recognition~\cite{metaBiasingList, BrianTCPGen,CPPBiasing, SelfAttnBiasing}, namely contextualized ASR.
Depending on the form of the context, most existing contextualized ASR systems fall into two categories: word-level context and utterance-level context. Word-level context biasing aim to improve the recognition accuracy of rare words such as contact names or application names. 
Sun et.al~\cite{BrianTCPGen} used a tree-constrained pointer generator to boost the posterior of words in a context list if the prefix matches during decoding. Huang et.al~\cite{CPPBiasing} proposed a neural network-based method to improve rare-word recognition on various E2E ASR architectures with a context phrase prediction network. 
Unlike word-level context, utterance-level context carries more sophisticated information such as topic and logical relationships. Text embeddings encoded by BERT~\cite{BERT} are utilized ~\cite{wei2021attentive} to improve ASR performance in multi-turn dialogues. Similarly, Chang et. al~\cite{SelfAttnBiasing} improve long-form ASR~\cite{longformASR} performance on neural transducer ~\cite{RNNT} using self-attentive embeddings from BERT. Li et.al~\cite{llamaPrompt} used LLaMa~\cite{llama} as the decoder of a speech encoder to facilitate domain adaptation through text prompts such as titles and topic descriptions.

In large language models, prompts are crucial to the correctness, fluency and format of the generated text~\cite{PromptTuning, PromptEngineering}. Inspired by this, we propose a novel E2E ASR framework named PromptASR, which utilizes text prompts for contextualized speech recognition. In specific, a dedicated text encoder is added to the E2E ASR system to ingest two types of prompt: content prompt and style prompt. The prior provides contextual information and the latter specifies the style of desired ASR transcriptions (e.g. casing and punctuation). The encoded prompts are injected to the ASR encoder via cross-attention with hidden speech representations. Unlike most existing approaches for contextualized ASR that are specialized for either word or utterance-level context, PromptASR is able to benefit from both of them.
When decoded with the ground truth preceding text as content prompt, PromptASR achieves 21.9\% and 6.3\% relative word-error-rate (WER) reduction compared to a baseline ASR system on a book reading dataset and an in-house dataset. On a word-level context biasing task~\cite{metaBiasingList}, PromptASR achieves 13.4\% relative WER reduction even with biasing lists containing 1000 distractors. Finally, we show that the style prompt effectively guides the style of transcriptions. It is noteworthy that Whisper~\cite{Whisper} mentioned having a similar prompting mechanism, but the details are not included in their paper.
\vspace{-0.7em}



\vspace{-2mm}
\section{PromptASR}
\label{sec:method}
\vspace{-0.6em}
\subsection{System Architecture}
\begin{figure}[h]
    \centering
    \includegraphics[width=0.82\linewidth]{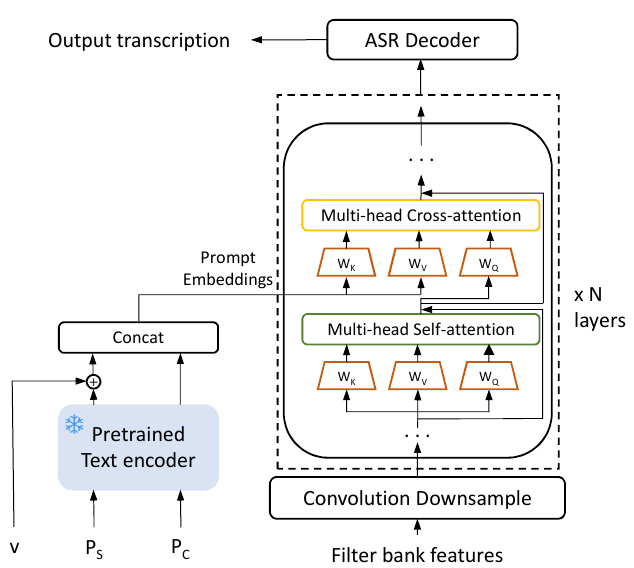}
    \caption{The architecture of PromptASR. The module in the dashed block is a transformer-like layer with cross-attention (other modules omitted). The text embeddings are injected as key/value pairs in the cross-attention module.}
    \label{fig:pipeline}
\end{figure}

The architecture of PromptASR is illustrated in Fig.~\ref{fig:pipeline}. It consists of a pre-trained text encoder $\textit{Enc}^{T}$, a speech encoder $\textit{Enc}^A$ and an ASR decoder $\textit{Dec}^A$. The text encoder $\textit{Enc}^{T}$ processes prompts and generates text embeddings. $\textit{Enc}^A$ consists of N transformer-like layers, each with a cross-attention module (with residual connection) placed after the self-attention module. Each layer receives not only the acoustic embeddings, but the text embeddings encoded by $\textit{Enc}^{T}$. The fusion between the text modality and speech modality is achieved by cross attention, where the text embeddings serve as key/value pairs and acoustic hidden states serve as query. The whole system can be trained with any ASR objective functions. 
\vspace{-3mm}
\vspace{-4mm}

\subsection{Prompts}
\vspace{-2mm}

Two types of prompts are defined in PromptASR: content prompt and style prompt. 
Content prompt should contain semantic and context-related information, which are usually in the form of sentences or a list of rare-words to be boosted. 
Most existing ASR systems produce normalized transcriptions, requiring inverse text normalization for production scenarios. Motivated by this, we would like the model to output different styles of transcription given different style prompts. The style prompt should indicate the desired style of the ASR transcription, such as casing and punctuation. During training, the style of the target text should always match the style prompt. Two training samples of PromptASR with different styles are shown in Table~\ref{tab:style prompt}.

\begin{table}[h!]
\adjustbox{max width=0.95\linewidth}{
\begin{tabularx}{\linewidth}{|l|X|}
\hline
Style Prompt   & \small{WITHOUT CASING OR PUNCTUATION}                   \\ \hline
Content Prompt & Welcome to the UEFA Champions League final!        \\ \hline
Reference text & \small{TODAY'S MATCH IS BETWEEN REAL MADRID AND LIVERPOOL} \\ \hline
\hline
\vspace{0.1em}
Style Prompt   & Mixed-cased English with punctuation              \\ \hline
Content Prompt & Welcome to the UEFA Champions League final!        \\ \hline
Reference text & Today's match is between Real Madrid and Liverpool. \\ \hline
\end{tabularx}}
\vspace{-1mm}
\caption{Two training samples with different style prompts.}
\label{tab:style prompt}
\vspace{-1em}
\end{table}

The tokenized content prompt $\textbf{P}_c=\textbf{P}_{c,1},\cdots,\textbf{P}_{c,n}$ and style prompt $\textbf{P}_s=\textbf{P}_{s,1},\cdots,\textbf{P}_{s,m}$ are feed to the text encoder to produce prompt embeddings $\mathcal{E}_c \in \mathbb{R}^{n\times c}$ and $\mathcal{E}_s \in \mathbb{R}^{m\times c}$. To distinguish between content prompts and style prompts, a trainable style indicator vector $v \in \mathbb{R}^c$ is added to the embeddings of the style prompts. The forward process of PromptASR is formulated in Eqn~\ref{eqn:prompt asr}.

\vspace{-1.5em}
\begin{align}
\begin{split}
    \mathcal{E}_c &= \textit{Enc}^T (\textbf{P}_c); \\
    \mathcal{E}_s &= \textit{Enc}^T (\textbf{P}_s) + v; \\
    \mathcal{G} &= \textit{Enc}^A(\textbf{Concat}(\mathcal{E}_c, \mathcal{E}_s), \mathbf{X}); \\ 
    \textbf{y} &= \textit{Dec}^A(\mathcal{G}),
\end{split}
\label{eqn:prompt asr}
\end{align}
\vspace{-0.1em}
where $\textbf{X}$ is the input speech features and $\textbf{y}$ is the output transcription. $\textbf{Concat}$ is the operation of concatenating two tensors along time axis.

To deal with situations where prompts are unavailable, both prompts are dropped out by a small probability during training so that the model learns to transcribe without prompts. In real-life scenarios, the content prompts can be completely irrelevant to the current utterance, requiring the model to learn to ignore such prompts. Therefore, a small proportion of the content prompts within the mini-batch are exchanged to simulate this scenario for better robustness to irrelevant prompts. 


\vspace{-3mm}
\section{Experiment Setup}
\label{sec:experiment}
\vspace{-2mm}
\subsection{Dataset}
\vspace{-1.5mm}
The open-sourced dataset Libriheavy~\cite{Libriheavy} is chosen as one of the training sets of PromptASR, as each utterance in Libriheavy is provided with a ground truth transcription and its preceding text of length 1000 bytes. Casing and punctuation of both texts are preserved. The medium subset containing around 5000 hours transcribed book readings is adopted in this work. The official Libriheavy test-clean and test-other sets are adopted for evaluation, which are approximately ~20\% harder (in terms of word error rates) than the LibriSpeech~\cite{Librispeech} test-clean and test-other sets. Additional 2000 hours recordings of conversations and podcasts covering different topics are also collected from the National Public Radio (NPR). An official transcript with casing and punctuation is provided for each recording. The text-audio alignments are obtained based on the Levenshtein distance between the output of a pre-trained ASR system and the official transcript and verified by human experts. The 1000-byte-long preceding text for each utterance is also extracted according to the alignment. 18 hours of recordings are hold-out and form the NPR evaluation set. 
\vspace{-3.2mm}
\subsection{Model Selection}
\vspace{-0.3em}

The pre-trained BERT~\cite{BERT} model is selected as the text encoder as it captures contextualized information through the masked language modeling pre-training. In addition, we also pre-train two transformers using the BERT objective on the text data of Libriheavy or NPR. Note that there are no overlaps between the training and testing books/recordings. The parameters of the text encoder are frozen during training.

The ASR system is a neural transducer with a Zipformer~\cite{Zipformer} speech encoder. A stateless decoder~\cite{StatelessRNNT} seeing two previous tokens and a joint network is added and pruned-RNNT~\cite{prunedRNNT} loss is used as the training objective. Chunk-wise streaming~\cite{Zipformer} is adopted for training streaming ASR models.

\vspace{-4mm}
\subsection{PromptASR Training}
\label{sec:prompts}
\vspace{-1.5mm}

Each utterance's preceding 1000 bytes are used as the content prompt during training. Two types of style transforms are pre-defined: \textbf{U}pper-\textbf{C}ased without punctuation (UC) and \textbf{M}ixed-\textbf{C}ased with \textbf{P}unctuation (MCP). A style is sampled for each utterance in the mini-batch and is applied to its style prompt and reference prompt. The \textbf{MCP} has a higher sampling probability (0.7) since it is more production-friendly. The style prompt is a sub-string of content prompt from other samples in the same mini-batch. 
To construct word-list based content prompt, the number of total appearances of each word in the training set is counted and the words outside the most common 10000 words are regarded as rare words. A context list is formed for each training sample by picking up rare words present in this sample and adding 50-100 randomly distractors. Together with the preceding text, both types of content prompts are used by probability of 0.5.

During training, SpecAug~\cite{specaug} and MUSAN~\cite{musan} are adopted to augment the training data. Speed perturbation is not used. The 80-D mel filter bank features are used as the input acoustic features. Byte-pair encoding~\cite{BPE} with byte fallback is used as modelling units and the vocabulary size is 500. The model is trained for 50 epochs on Libriheavy medium subset, and 60 epochs on the NPR dataset. The checkpoints of last ten epochs are averaged and beam search of size 4 is used for inference.

\vspace{-4mm}
\subsection{Metrics}
\vspace{-0.4mm}
Two types of evaluation scenarios are investigated. First, each test sample is given its ground truth preceding 1000 bytes as content prompt during decoding. This can be seen as the performance upper bound of the PromptASR model when dealing with utterance-level content prompt. However, the ground truth preceding text is not always available in real-life scenarios and the model has to rely on the erroneous decoding results of the previous utterances. Therefore, a 15-hour long-form recordings test set is also collected from NPR to validate the long-form ASR performance. The long recordings are split into individual sentences without overlap. The average length of the recording is 20 minutes. The decoding results from previous sentences are concatenated to construct the content prompt for the current utterance and a fixed style prompt irrelevant to the recording is used.

To test the capability of PromptASR on word-level context biasing, the biasing list for LibriSpeech~\cite{Librispeech} test sets in ~\cite{metaBiasingList} is used. Each utterance in the test set is provided with a biasing word list containing biasing words and distractors. Note that the biasing lists of some utterances are purely distractors. 

\vspace{-2mm}
\section{Experiment Results}
\vspace{-2mm}

\subsection{Utterance-level Context Biasing}
\vspace{-1mm}
Experiments are first carried out to validate the benefit of content prompts, where the ground truth preceding text is used as content prompt during decoding. Baseline neural transducer models without text encoder are trained with UC and MCP styles separately. The word-error-rates (WERs) are shown in Table~\ref{tab:WERs} and the following points can be drawn.
First, PromptASR model significantly improves the WERs owing to the contextual information from the content prompts. For non-streaming models, relative WER reductions (WERRs) of 21.9\% and 6.8\% are achieved compared to the baseline (\textbf{B1-UC}) on the Libriheavy test-other and NPR with the in-domain text encoder using style UC. Similar relative WERRs of 20.0\% and 7.8\% are observed for streaming models. If no content prompt is given, PromptASR still achieves comparable results as the baseline model, suggesting that the model is robust to decode without any prompts.
Second, pre-training the text encoder on in-domain data further improves the performance on the in-domain test sets. 
Finally, the style change in PromptASR does not affect the WER. After normalizing the transcript with UC style, decoding with MCP style yields similar WERs as using UC style.

\vspace{-1.5mm}
\begin{table}[h!]
    \centering
    \adjustbox{max width=\linewidth}{
    \begin{tabular}{lccccc}
    \toprule
      \multirow{2}{*}{Model}   & \multirow{2}{*}{\shortstack[c]{Content\\Prompt}} & \multirow{2}{*}{\shortstack[c]{Style\\Prompt}} & \multicolumn{2}{c}{Libriheavy} & \multirow{2}{*}{NPR} \\
      \cmidrule(lr){4-5}
      & & & clean & other & \\
      \midrule
      \midrule
      \multicolumn{3}{l}{\textbf{Non-streaming Models}} \\
      \textbf{B1-UC}   &  - & - & 3.0  & 6.72 &  5.09   \\
      \textbf{B1-MCP}  &  - & - & 3.11 (10.4) & 6.74 (14.3) &  5.26 (9.5)  \\
        
      \midrule
      \multirow{3}{*}{\shortstack[c]{BERT~\cite{BERT}\\Text Enc}}   &  - & - &  3.14 & 6.71  & 5.26  \\
         &  \checkmark & UC  & 2.82  &  6.03  & 4.89 \\
         &  \checkmark & MCP & 2.64 (9.3)   &  5.55 (12.5)   & 4.95 (8.48) \\
         
      \midrule
      \multirow{3}{*}{\shortstack[c]{In-domain\\Text Enc}} & - & - & 3.09 &  6.8 & 5.28 \\ 
         &  \checkmark & UC  & 2.48 & 5.25 & 4.74 \\
         &  \checkmark & MCP & 2.38 (9.0) & 5.03 (12.2) & 4.8 (8.4) \\
         
      \midrule
      \midrule
      \multicolumn{3}{l}{\textbf{Streaming Models}} \\
      \textbf{B2-UC} &  - & - & 3.73  & 8.19 & 6.77 \\
      \midrule
      \multirow{3}{*}{\shortstack[c]{In-domain\\Text Enc}} & - & -  & 3.7 & 8.13 & 6.96  \\
        & \checkmark  & UC  & 3.01 & 6.55 & 6.24 \\
        & \checkmark  & MCP & 2.93 (10.5) & 6.29 (14.4) & 6.29 (10.8)  \\
    \bottomrule
    \end{tabular}}
    \caption{WERs (\%) of baseline models and PromptASR models on different test sets. For MCP style, WERs before (in brackets) and after UC style text normalization are reported.}
    \vspace{-0.3em}
    \label{tab:WERs}
\end{table}

\vspace{-3mm}
\FloatBarrier
\begin{table*}[!htbp]
\centering
\begin{tabular}{|l|cccccccccc|cc|}
\hline
\multirow{3}{*}{Model} & \multicolumn{10}{c|}{Word-level LibriSpeech Biasing}                                                                                                                                    & \multicolumn{2}{l|}{\multirow{2}{*}{\shortstack[c]{Utterance-level\\Libriheavy Biasing}}} \\ \cline{2-11}
                       & \multicolumn{2}{c|}{No Biasing}                     & \multicolumn{2}{c|}{N=0}                                               & \multicolumn{2}{c|}{N=100}                                              & \multicolumn{2}{c|}{N=500}                                             & \multicolumn{2}{c|}{N=1000}                                     & \multicolumn{2}{l|}{}                            \\ \cline{2-13} 
                       & \multicolumn{1}{c|}{clean} & \multicolumn{1}{c|}{other} & \multicolumn{1}{c|}{clean}         & \multicolumn{1}{c|}{other}        & \multicolumn{1}{c|}{clean}         & \multicolumn{1}{c|}{other}         & \multicolumn{1}{c|}{clean}        & \multicolumn{1}{c|}{other}         & \multicolumn{1}{c|}{clean}         & \multicolumn{1}{c|}{other} & \multicolumn{1}{c|}{\;\;clean\;\;}        & \multicolumn{1}{c|}{other}        \\ \hline
\textbf{B1-UC}            & \multicolumn{1}{c|}{2.46}  & \multicolumn{1}{c|}{5.11}  & \multicolumn{1}{c|}{2.46}          & \multicolumn{1}{c|}{5.11}         & \multicolumn{1}{c|}{2.46}          & \multicolumn{1}{c|}{5.11}          & \multicolumn{1}{c|}{2.46}         & \multicolumn{1}{c|}{5.11}          & \multicolumn{1}{c|}{2.46}          & 5.11                       & \multicolumn{1}{c|}{3.0}         & 6.72         \\ \hline
\textbf{M1}                     & \multicolumn{1}{c|}{2.45}  & \multicolumn{1}{c|}{5.09}  & \multicolumn{1}{c|}{2.37}          & \multicolumn{1}{c|}{4.90}         & \multicolumn{1}{c|}{2.49}          & \multicolumn{1}{c|}{5.36}          & \multicolumn{1}{c|}{2.62}             & \multicolumn{1}{c|}{5.71}              & \multicolumn{1}{c|}{2.69}              &     \multicolumn{1}{c|}{5.82}             & \multicolumn{1}{c|}{2.48}         & 5.25         \\ \hline
\textbf{M2}                     & \multicolumn{1}{c|}{2.43}  & \multicolumn{1}{c|}{5.07}  & \multicolumn{1}{c|}{\textbf{1.73}} & \multicolumn{1}{c|}{\textbf{4.0}} & \multicolumn{1}{c|}{\textbf{1.73}} & \multicolumn{1}{c|}{\textbf{4.07}} & \multicolumn{1}{c|}{\textbf{2.0}} & \multicolumn{1}{c|}{\textbf{4.45}} & \multicolumn{1}{c|}{\textbf{2.13}} & \textbf{4.67}              & \multicolumn{1}{c|}{2.59}         & 5.55          \\ \hline
\end{tabular}
\vspace{-1mm}
\caption{WERs (\%) on the word-level LibriSpeech biasing task of the baseline model and PromptASR model with different sizes of biasing list. $N$ is the number of distractors added to the biasing word list. The WERs of the Libriheavy utterance-level biasing test are also shown for reference.}
\label{tab:LS biasing}
\vspace{-1.5em}
\end{table*}

\vspace{-4mm}
\subsection{Long-form ASR}
\vspace{-1mm}

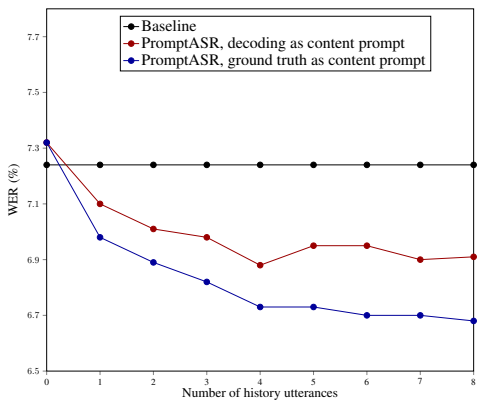
\begin{figure}[h]
    \centering
    \begin{tikzpicture}[scale=0.35]
\definecolor{red}{RGB}{154,0,0}
\definecolor{blue}{RGB}{0,0,154}

\begin{axis}[
    legend cell align={left},
    legend style={
        nodes={scale=1.2, transform shape},
        at={(0.9, 0.98)},
        font=\Large
    },
    label style={font=\Large},
    width=\textwidth,
    tick align=outside,
    tick pos=left,
    xlabel=Number of history utterances,
    ylabel= WER (\%),
    xmin=0, xmax=8,
    ymin=6.5, ymax=7.8,
    xtick={0,1,...,8},
    xticklabels={0,1,...,8},   
    ytick={6.5,6.7,...,7.8}
            ]
\addplot[mark=*,color=black, mark size=3,black] 
plot coordinates {
        (0,7.24)
        (1,7.24)
        (2,7.24)
        (3,7.24)
        (4,7.24)
        (5,7.24)
        (6,7.24)
        (7,7.24)
        (8,7.24)
};
\addlegendentry{Baseline}

\addplot[color=red,mark=*, mark size=3, red]
    plot coordinates {
        (0,7.32) 
        (1,7.1)
        (2,7.01)
        (3,6.98)
        (4,6.88)
        (5,6.95)
        (6,6.95)
        (7,6.90)
        (8,6.91)
    };
\addlegendentry{PromptASR, decoding as content prompt}

\addplot[color=blue,mark=*, mark size=3, blue]
    plot coordinates {
        (0,7.32)
        (1,6.98)
        (2,6.89)
        (3,6.82)
        (4,6.73)
        (5,6.73)
        (6,6.70)
        (7,6.70)
        (8,6.68)
    };
\addlegendentry{PromptASR, ground truth as content prompt}

\end{axis}
\end{tikzpicture}
    \vspace{-2mm}
    \caption{WERs (\%) of baseline model and PromptASR models after normalization on the long-form ASR task.}
    \label{fig:long form}
    \vspace{-3mm}
\end{figure}

The performance of PromptASR on long-form ASR is investigated and results are shown in Fig~\ref{fig:long form}. The non-streaming PromptASR model with in-domain text encoder trained on NPR is decoded in UC style with either erroneous decoding results (red) or the ground truth transcripts (blue) of the history utterances as content prompt. For reference, the WER of a baseline ASR system \textbf{B1-UC} from Table~\ref{tab:WERs} is also plotted (black). Though both content prompts reduce the WER compared to the baseline, the gain from using erroneous decoding results of preceding utterances is smaller and it fails to further improve the WER with a history of longer than 4 utterances. This could be caused by the wrong prediction of named entities or keywords, which undermines the benefit from context information for PromptASR.
\vspace{-1.3mm}
\vspace{-2mm}

\vspace{-3.5mm}
\subsection{Word-level Context Biasing}
\vspace{-1mm}

The potential of applying PromptASR for rare-word recognition is investigated. All models in Table~\ref{tab:LS biasing} are trained on Libriheavy medium subset. Note that only the in-domain text encoder can be used for this task, as the official release of BERT has a maximum input length constraint of 512 tokens.
\textbf{B1-UC} is the non-streaming baseline model from Table~\ref{tab:WERs}. \textbf{M1} and \textbf{M2} are PromptASR models sharing the same text encoder pre-trained on the Libriheavy large subset (disjoint from LibriSpeech test sets). \textbf{M1} only uses the previous utterances as content prompt during training, while \textbf{M2} additionally uses content prompts constructed from rare-words list as described in Sec~\ref{sec:prompts}. 
The WERs with different sizes of context lists on the LibriSpeech biasing task are shown in Table~\ref{tab:LS biasing}. All three models yield similar performance without biasing lists. When decoded with biasing lists, \textbf{M1} fails to benefit from the word-level context. However, only at the cost of slight WER degradation on the utterance-level context biasing (last two columns), constructing content prompts from rare-words list (\textbf{M2}) during training significantly helps word-level contextual biasing, achieving relative WERRs of 29.7\% and 21.7\% at $N=100$ compared to the baseline model. However, as the number of distractors increases, the performance gain decreases very quickly and the relative WERRs at $N=1000$ are 13.4\% and 8.6\%. One possible reason is that the context list used during training is shorter than 100 words, thus the model did not generalize well to larger $N$. 
Despite this, the relative improvements are still comparable with the existing neural network-based context biasing methods~\cite{CPPBiasing}, which yields 10.6\% and 12.6\% relative WERRs with $N=1000$.


\vspace{-4.3mm}
\subsection{Output Examples}
\vspace{-0.4em}
A few examples of the output of PromptASR model is shown below. In Table~\ref{tab:example output}, the first block shows an example where the model outputs transcriptions with accurate casing and punctuations using MCP style. The second block shows that the PromptASR model corrects the ASR output with the help of content prompt -- it learns from the content prompt that the topic is about ``horse'', and guides the model to output the in-domain word ``phaeton'' instead of a made-up word ``faithon''.
\vspace{-4mm}
\begin{table}[H]
\adjustbox{max width=1.0\linewidth}{
\begin{tabularx}{\linewidth}{|l|X|}
\hline
style prompt & Mixed-cased English with punctuations. \\ \hline
\multirow{2}{*}{\shortstack[l]{PromptASR \\output}} & ``Do you believe in some education?'' asked Mary Taylor.\\ \hline
\hline
\multirow{2}{*}{\shortstack[l]{content \\prompt}} & ... I knew how hard it was upon slow-paced \textbf{horses} to be put with fast ones; ...  \\ \hline
\multirow{2}{*}{\shortstack[l]{Baseline \\output}}       & She was often used in the \textcolor{red}{faithon}, and was very much liked by some of the ladies. \\ \hline
\multirow{2}{*}{\shortstack[l]{PromptASR \\output}}      & She was often used in the \textcolor{blue}{phaeton}, and was very much liked by some of the ladies. \\ \hline
\end{tabularx}}

    \caption{Outputs of PromptASR versus normal ASR system.}
    \label{tab:example output}
    \vspace{-4mm}
\end{table}
\vspace{-5mm}
\section{Conclusion}
\vspace{-2.3mm}
\label{sec:conclusion}
We propose the PromptASR framework, which performs contextualized ASR with controllable style of transcriptions. By passing either the ground truth transcript or decoded transcript of previous utterances as content prompt, the PromptASR model utilizes the cross-utterance context and improves the WER compared to a baseline ASR system. If the content prompt is a list of biasing words, the PromptASR model can also perform word-level biasing and achieve significant WER reduction on biasing words while being robust to distractors. The model can switch the output style (e.g. casing and punctuation) given different style prompts. In the future, we hope to explore more efficient utilization of text embeddings to reduce the computational cost. We are also very interested in how to incorporate large language models into PromptASR.




\vfill\pagebreak

\bibliographystyle{IEEEbib}
\bibliography{refs}

\begin{thebibliography}{10}

\bibitem{ASRrescoring}
Richard Schwartz and Steve Austin,
\newblock ``A comparison of several approximate algorithms for finding multiple
  ({N}-best) sentence hypotheses,''
\newblock in {\em Proc. ICASSP}, Toronto, 1991.

\bibitem{TonyGPT}
Xianrui Zheng, Chao Zhang, and Philip~C Woodland,
\newblock ``Adapting {GPT}, {GPT-2} and {BERT} language models for speech
  recognition,''
\newblock in {\em Proc. ASRU}, Cartagena, 2021.

\bibitem{shallowfusion}
Anjuli Kannan, Yonghui Wu, Patrick Nguyen, Tara~N Sainath, Zhijeng Chen, and
  Rohit Prabhavalkar,
\newblock ``An analysis of incorporating an external language model into a
  sequence-to-sequence model,''
\newblock in {\em Proc. ICASSP}, Calgary, 2018.

\bibitem{metaBiasingList}
Duc Le, Mahaveer Jain, Gil Keren, Suyoun Kim, Yangyang Shi, Jay Mahadeokar,
  Julian Chan, Yuan Shangguan, et~al.,
\newblock ``Contextualized streaming end-to-end speech recognition with
  trie-based deep biasing and shallow fusion,''
\newblock in {\em Proc. Interspeech}, Brno, 2021.

\bibitem{BrianTCPGen}
Guangzhi Sun, Chao Zhang, and Philip~C Woodland,
\newblock ``Tree-constrained pointer generator for end-to-end contextual speech
  recognition,''
\newblock in {\em Proc. ASRU}, Cartagena, 2021.

\bibitem{CPPBiasing}
Kaixun Huang, Ao~Zhang, Zhanheng Yang, Pengcheng Guo, et~al.,
\newblock ``Contextualized end-to-end speech recognition with contextual phrase
  prediction network,''
\newblock in {\em Proc. Interspeech}, Dublin, 2023.

\bibitem{SelfAttnBiasing}
Shuo-Yiin Chang, Chao Zhang, Tara~N Sainath, Bo~Li, and Trevor Strohman,
\newblock ``Context-aware end-to-end {ASR} using self-attentive embedding and
  tensor fusion,''
\newblock in {\em Proc. ICASSP}, Rhodes, 2023.

\bibitem{BERT}
Jacob Devlin, Ming-Wei Chang, Kenton Lee, and Kristina Toutanova,
\newblock ``Bert: Pre-training of deep bidirectional transformers for language
  understanding,''
\newblock in {\em Proc. NAACL}, Minneapolis, 2019.

\bibitem{wei2021attentive}
Kai Wei, Thanh Tran, Feng-Ju Chang, et~al.,
\newblock ``Attentive contextual carryover for multi-turn end-to-end spoken
  language understanding,''
\newblock in {\em Proc. ASRU}, Cartagena, 2021.

\bibitem{longformASR}
Arun Narayanan, Rohit Prabhavalkar, Chung-Cheng Chiu, David Rybach, Tara~N
  Sainath, and Trevor Strohman,
\newblock ``Recognizing long-form speech using streaming end-to-end models,''
\newblock in {\em Proc. ASRU}, 2019.

\bibitem{RNNT}
A.~Mohamed A.~Graves and G.~Hinton,
\newblock ``Speech recognition with deep recurrent neural networks,''
\newblock in {\em Proc. ICASSP}, Vancouver, 2013.

\bibitem{llamaPrompt}
Yuang Li, Yu~Wu, Jinyu Li, and Shujie Liu,
\newblock ``Prompting large language models for zero-shot domain adaptation in
  speech recognition,''
\newblock {\em arXiv preprint arXiv:2306.16007}, 2023.

\bibitem{llama}
Hugo Touvron, Louis Martin, Kevin Stone, Peter Albert, Amjad Almahairi, Yasmine
  Babaei, et~al.,
\newblock ``Llama 2: Open foundation and fine-tuned chat models,''
\newblock {\em arXiv preprint arXiv:2307.09288}, 2023.

\bibitem{PromptTuning}
Brian Lester, Rami Al-Rfou, and Noah Constant,
\newblock ``The power of scale for parameter-efficient prompt tuning,''
\newblock in {\em Proc. EMNLP}, Punta Cana, 2021.

\bibitem{PromptEngineering}
Laria Reynolds and Kyle McDonell,
\newblock ``Prompt programming for large language models: Beyond the few-shot
  paradigm,''
\newblock in {\em Proc. CHI}, New York, 2021.

\bibitem{Whisper}
Alec Radford, Jong~Wook Kim, Tao Xu, Greg Brockman, Christine McLeavey, and
  Ilya Sutskever,
\newblock ``Robust speech recognition via large-scale weak supervision,''
\newblock in {\em Proc. ICML}, Hawaii, 2023.

\bibitem{Libriheavy}
Wei Kang et~al.,
\newblock ,''
  \url{https://github.com/k2-fsa/text_search/tree/master/examples/libriheavy},
  2023.

\bibitem{Librispeech}
Vassil Panayotov, Guoguo Chen, Daniel Povey, and Sanjeev Khudanpur,
\newblock ``Librispeech: an {ASR} corpus based on public domain audio books,''
\newblock in {\em Proc. ICASSP}, Brisbane, 2015.

\bibitem{Zipformer}
Zengwei Yao, Liyong Guo, Xiaoyu Yang, Wei Kang, Fangjun Kuang, Yifan Yang,
  Zengrui Jin, Long Lin, and Daniel Povey,
\newblock ``Zipformer: A faster and better encoder for automatic speech
  recognition,''
\newblock {\em arXiv preprint arXiv:2310.11230}, 2023.

\bibitem{StatelessRNNT}
Mohammadreza Ghodsi, Xiaofeng Liu, James Apfel, Rodrigo Cabrera, and Eugene
  Weinstein,
\newblock ``{RNN}-transducer with stateless prediction network,''
\newblock in {\em Proc. ICASSP}, Barcelona, 2020.

\bibitem{prunedRNNT}
Fangjun Kuang, Liyong Guo, Wei Kang, Long Lin, Mingshuang Luo, Zengwei Yao, and
  Daniel Povey,
\newblock ``Pruned {RNN-T} for fast, memory-efficient {ASR} training,''
\newblock in {\em Proc. Interspeech}, Incheon, 2022.

\bibitem{specaug}
Daniel~S Park, William Chan, Yu~Zhang, Chung-Cheng Chiu, Barret Zoph, et~al.,
\newblock ``Specaugment: A simple data augmentation method for automatic speech
  recognition,''
\newblock in {\em Proc. Interspeech}, Graz, 2019.

\bibitem{musan}
David Snyder, Guoguo Chen, and Daniel Povey,
\newblock ``{MUSAN}: {A} {M}usic, {S}peech, and {N}oise {C}orpus,''
\newblock {\em arXiv:1510.08484v1}, 2015.

\bibitem{BPE}
Rico Sennrich, Barry Haddow, and Alexandra Birch,
\newblock ``Neural machine translation of rare words with subword units,''
\newblock in {\em Proc. ACL}, Berlin, 2016.

\end{thebibliography}

\end{document}